\documentclass[prl,aps,amssymb,amsmath,superscriptaddress,twocolumn,showpacs]{revtex4}

\usepackage{amsmath,amsfonts,amsthm,amssymb,amsxtra,dsfont}

\theoremstyle{definition}

\theoremstyle{remark}

\renewcommand{\k}{\mathbf{k}}

\renewcommand{\l}{\mathbf{l}}

\newcommand{\p}{\mathbf{p}}
\newcommand{\q}{\mathbf{q}}

\newcommand{\F}{\mathcal{F}}

\renewcommand{\phi}{\varphi}
\newcommand{\R}{\mathbb{R}}

\DeclareMathOperator{\sgn}{sgn}
\DeclareMathOperator{\Tr}{Tr}

\begin{document}

\title{General pairing mechanisms in the BCS-theory of superconductivity}

\author{Christian Hainzl} 
\affiliation{Mathematisches Institut, Universit\"at T\"ubingen, Auf der Morgenstelle 10, 72076 T\"ubingen, Germany}
\author{Michael Loss}
\affiliation{School of Mathematics, Georgia Tech, Atlanta GA 30332}

\begin{abstract}
Starting from the linearized BdG-equation we make the simple observation that pairing can occur between particles with total momenta different from zero, e.g., with equal momentum and opposite spin, in cases of an effective interaction acting in the center-of-mass coordinates. 
\end{abstract}

\pacs{x}

\maketitle

Since the ground-breaking work \cite{bcs} of Bardeen, Cooper, and Schrieffer in 1957, it is well established that the primary cause for superconductivity 
in classical metals is the formation and condensation of Cooper-pairs. These pairs are formed via a small effective interaction mediated by lattice vibrations, i.e., phonons. Once these pairs form, they are macroscopically coherent and lead to off diagonal long range order (ODLRO). 
In cuprates, however, where the superconductive state depends significantly on the chemical stoichiometry of the material, this mechanism is much less established. Nonetheless, it is widely accepted, e.g. \cite{leggett_quantum_liquids},
 that superconductivity is still due to the formation of pairs and these pairs are the primary carriers of ODLRO, but the eventual cause for the effective interaction, not to mention its eventual form, 
is still highly disputed even 30 years after its initial discovery \cite{BM}. 

Thus, as our starting point we assume  pairs to be the primary cause for superconductivity and consider the two-particle BdG (Bogolubov-de-Gennes) equation \cite{degennes}. More precisely, since the superconductive phase transition is accepted to be 
of second order, we simplify matters by starting with the \emph{linearized BdG equation}. We allow a general two-body interaction $V=V\left(x-y, \frac {x+y}2\right)$, which depends on the relative
as well as COM (center-of-mass) coordinates $\frac {x+y}2$. We imagine the possibility of rather complicated effective interactions between electrons, however, it is not obvious how such an interaction could be mediated by the environment. 
One possible way might the following. Imagine a cuprate-type crystal with copper-oxide planes. If one electron causes a copper atom to vibrate such that these vibrations propagate to the neighboring copper atoms via their copper-oxide bonds, then these propagating vibrations 
further influence other electrons.
This might lead to an effective interaction which does not only depend on the relative distance between two electrons but also on their COM-variables. Observe, such an interaction would be 
periodic in the COM-variable.
The primary source of such an interaction would be of phonon-type, but strongly depending on the chemical bonding between the corresponding atoms and therefore cannot be related to the isotope-effect.

Our main message relies on the following simple observation:
The resolvent of the temperature dependent two-body dispersion relation \eqref{matsubara} 
has two types of singularities in the limit where the temperature goes to zero, $T \to 0$. The familiar one where the total momentum of the pairs vanish, $\p + \q =0$,
and the relative momentum $\k = (\p-\q)/2$ is close to the Fermi-momentum $k_F$, $|\k| \simeq k_F.$ This leads to the classical  BCS-equation for the critical temperature $T_c$.
However, there is another singularity for $\q = \p$, where the modulus of total momentum 
$\l = \p + \q$ is twice the Fermi momentum, $|\l| =  2 k_F$. If one  imagines systems with an effective potential 
acting primarily in the COM coordinates with a significant contribution in momentum space around 
twice the Fermi-momentum then this singularity may well describe the onset of pairing, where the pairs have a 
small relative momentum and a total momentum about $2 k_F$.  More  generally, the singularity of the resolvent takes place at a four dimensional manifold, with $|\p| = |\q|$, i.e., rather arbitrary pairing mechanisms are possible for general two-body  interactions. 
However, the effective potential has to be sharply concentrated on this four dimensional manifold in order to guarantee binding.
Once concentrated, binding does not depend on the details of the potential. It depends mainly on the singularity of the resolvent. 
For general bounded potentials supported in the $\p,\q$-space the corresponding integral may be finite in the $T \to 0$ limit, and 
 the existence of a solution depends sensitively on the corresponding parameters. 

From our basic assumptions we will draw the following conclusions:
First, if one considers an effective interaction with main contribution in the COM-coordinates pairing can occur for particles with 
equal momentum and opposite spin. If this effective potential has a significant contribution around $|\l|\simeq 2 k_F$ then
the associated critical temperature for the onset of pairing can be significantly increased compared to the situation of classical 
superconductors.  
Second, such \emph{equal momentum pairing} may explain the strong spatial variations of the electron density and the strong inhomogeneity of the system \cite{Pan}.  
If we further assume that phonons might create an effective potential which is periodic in the COM-variables then 
the Cooper-pair wavefunction might reflect this periodicity and even exhibit PDW (pair density wave) and CDW (charge density wave) orders.
Third, if the potential $V$ is {\em not} accidentally located on the singular manifold, we point out that solutions of the BdG-equation depend very sensitive on the effective potential $V$.  Slight changes of $V$, 
or of the Fermi-momentum $k_F$, e.g., by means of hole doping, can therefore change the critical temperature dramatically or 
even suppress superconductivity as a whole. These conclusions seem to be in accordance with experimental observations. 

The appearance of pairing with non-vanishing center-of-mass momentum was suggested in the sixties by Fulde-Ferell \cite{FF} and by Larkin-Ovchinnikov \cite{LO1, LO2} independently and is nowadays referred to as FFLO phases. 
However, let us remark that the pairs we suggest are of totally different type. In our approach the relative and COM-momenta are set on equal footing and in the extreme case of the example \eqref{highttcT} the COM-behavior predominantly determines the critical temperature.

Although we use the Sommerfeld model to describe the electron motion, the statement qualitatively stays unchanged in more general cases,
e.g., for Bloch-band-type electrons. Further the actual shape of the Fermi-surface does not play any significant role \cite{HSMZ}.

Let us mention that in previous years considerable effort has been put into the mathematical analysis of the BCS-functional, e.g., \cite{HHSS,FHNS,HS-T_C, FHSS, FHSS2,FHSchS,HSey},  for rather general pair interactions $V$. For a comprehensive review see \cite{HSR}. For translation-invariant systems it was rigorously proven that, in the weak coupling limit, the critical temperature \cite{FHNS,HS-T_C}, as well as the gap, depends only on the lowest eigenvalue of an appropriate operator depending on the behavior of $V$ on the Fermi-surface. In \cite{FHSS,FHSS2} slow spatial fluctuations were studied with a rigorous proof of the derivation of the Ginzburg-Landau equation 
of Gorkov' \cite{Gorkov}. In this context the linearized Bogolubov-de-Gennes (BdG) equation was used to study the influence of constant magnetic fields \cite{FHLa}.

We use units such that the mass of the particles, preferably electrons, is $1/2$, $\hbar =1$ and $k_B=1.$

Let us denote the expectation value of a pair of particles as 
\begin{equation}\label{alph} \langle a_{\q,\sigma} a_{\p,\nu} \rangle = (i{\bf \sigma}_y)_{\nu,\sigma} \alpha(\p,\q),\end{equation}
where $\p$ and $\q$  are the corresponding momenta, assume 
the states to be {\em spin-singlets}. We further assume the particles to interact via a general two-particle interaction $V$,
which we write in the general form
$$ V = V(\p,\q; \p' \q').$$ 

We denote, as usually, the gap-function as $\Delta =  V \alpha$, i.e.,
$$\Delta(\p,\q) = \int_{\R^3} \int_{\R^3} \frac {d^3 \p'}{(2\pi)^3} \frac{d^3 \q'}{(2\pi)^3} V(\p,\q; \p' \q') \,  \alpha(\p',\q') .$$

We assume our system to be three dimensional, however, the following arguments also work in two dimensions. 

Our object of interest is the {\em linearized} BdG (Bogolubov-de-Gennes) equation, e.g.,  \cite[Sec. 7]{degennes} with corresponding extension to
general two-body interactions. We present a more mathematical derivation in the appendix. 
Using momentum representation the linearized gap-equation for arbitrary interaction $V$  can be written in the form 
\begin{multline}\label{BdG}
\Delta(\p,\q) = - \frac 12 \int_{\R^3} \int_{\R^3} \frac {d^3 \p'}{(2\pi)^3} \frac{d^3 \q'}{(2\pi)^3} V(\p,\q; \p', \q') 
 \, \times \\  L_\beta(\p',\q') \Delta(\p',\q'),
\end{multline}
with 
\begin{equation}\label{matsubara}
 L_\beta (\p,\q) =  \frac{\tanh\left (\frac \beta 2 (\p^2 - k_F^2)\right)+ \tanh\left (\frac \beta 2 (\q^2 - k_F^2)\right)}{ \p^2 + \q^2  - 2 k_F^2},
 \end{equation}
where $k_F$ denotes the Fermi-momentum and $\beta = 1/T$ the inverse temperature. Observe that the latter expression has the familiar 
form 
$$
L_\beta (\p,\q) = 2T \sum_{\omega_n}  \frac 1{(\xi(\p) - i \omega_n)(\xi(\q) + i \omega_n)},
$$
with $\xi(\p) = \p^2 - k_F^2$, and $\omega_n= 2\pi T(n + \frac 12)$ are the Matsubara frequences. Equation \eqref{BdG} can be derived from a second quantized Hamiltonian with interaction $V$ and the Sommerfeld dispersion relation by reducing the system to quasi-free or BCS-type-states, see e.g., \cite{bls, HSR} for a mathematical derivation, and omitting the direct interaction energy and the Fock term. 

%

\medskip

In order to simplify matters and to more clearly convey our message,  we perform a standard simplification and restrict our attention to potentials $V$ which are rank one operators on the two particle space,
$$V(\p,\q; \p',\q') = - g |v(\p,\q)\rangle \langle v(\p',\q')|,$$
with $v(\p,\q)$ a square integrable function in $\R^3 \times \R^3$, and $g$ is the coupling strength of the potential. 
The following argument, however, applies to very general interactions $V$. 

In an abstract notation the corresponding BdG-equation has the form 
$$ \Delta = \frac g2 |v\rangle \langle v| L_\beta \Delta \rangle.$$ 
This shows that a potential solution $\Delta$ has to be proportional to $|v\rangle$,
$\Delta = c_\beta |v\rangle$, with $c_\beta$ a temperature-dependent constant, such that
$$ c_\beta |v\rangle = \frac g2 c_\beta |v\rangle \langle v| L_\beta \Delta \rangle,$$
which is equivalent to $$1 = \frac g2 \langle v| L_\beta | v\rangle ,$$ i.e.,
\begin{multline}\label{BdGrank1}
\frac 1g = \frac 12 \int_{\R^3} \int_{\R^3} \frac {d^3 \p}{(2\pi)^3} \frac{d^3 \q}{(2\pi)^3} |v(\p,\q)|^2 \times \\  \frac{\tanh\left (\frac \beta 2 (\p^2 - k_F^2)\right)+ \tanh\left (\frac \beta 2 (\q^2 - k_F^2)\right)}{ \p^2 + \q^2  - 2 k_F^2}.
\end{multline}
Using relative, $\k = (\p - \q)/2$ and center-of-mass coordinates, $\l = \p + \q$, this equation takes the form
\begin{multline}\label{BdGrank1-rel}
\frac 1g = \frac 12 \int_{\R^3} \int_{\R^3} \frac {d^3 \k}{(2\pi)^3} \frac{d^3 \l}{(2\pi)^3} |v(\k + \l/2,-\k + \l/2)|^2 \times \\   \frac{\tanh\left (\frac \beta 2 ((\k + \l/2)^2 - k_F^2)\right)+ 
\tanh\left (\frac \beta 2 ((\k - \l/2)^2 - k_F^2)\right)}{ 2\k^2 + \l^2/2  - 2 k_F^2}.
\end{multline}
Observe that for $T \to 0$  singularities appear at $\l =0$ and $|\k| = k_F$, as well at 
$\k=0$ and $|\l| = 2 k_F$. If $V$ is supported in these areas then equation \eqref{BdGrank1-rel} is guaranteed to have a solution.
More general the singularity occurs at the four dimensional manifold $\{(\p,\q) \, | \, |\p| = |\q| = k_F\}$, and 
a solution to the pair equation is guaranteed for any $v(\p,\q)$, appropriately located on this surface. Observe that in terms 
$\l$ and $\k$ the singularity manifold is characterized by the relation $\l \cdot \k =0$. 
For potentials with bounded functions $v(\p,\q)$ the corresponding integral is always finite, and hence, does not 
in general guarantee a solution. In such situations the occurrence of pairing strongly depends on the involved parameters. 

\subsection{Classical superconductors}

In the case of classical superconductors the phonon-induced interaction is modeled 
by the effective interaction $$  |v(\k + \l/2,-\k + \l/2)|^2 = \theta(| |\k|^2 - k_F^2 | \leq \omega_D) \delta(\l),$$
where $\omega_D$ is the Debye-frequency and $\theta$ the characteristic function. 
Setting $\xi = |\k|^2 - k_F^2$ and $$\int_{\R^3} \frac {d^3\k}{(2\pi)^3} \simeq N(0) \int d\xi, $$
with $N(\xi) = \frac{\sqrt{\xi + k_F^2}}{(2\pi)^2}$ denoting the density of states,
one recovers the familiar equation for the critical temperature
$$ 1 = g N(0) \int_0^{\omega_D} \frac {d\xi}{\xi} \tanh \frac {\xi}{2T_c},$$ 
with approximate solution $T_c = \frac {2 e^{\gamma}}{\pi} \omega_D e ^{-1/N(0) g}$. 

%

\subsection{Pairs composed of $\p , \uparrow$ and $\p, \downarrow$}
 
In the case of highly anisotropic media, depending significantly on the stoichiometry of the system, such as cuprates in high-Tc superconductors, it may be possible to obtain an effective potential, which is a function of the COM coordinates as well. 
In this context we suggest a much more general mechanism for the onset of pairs   
than simply the classical one. 

For the sake of simplicity, let us take the extreme example of
$V$ only acting in the center of mass coordinates and assume an effective potential
 of the type $V = - \bar g | v\rangle \langle v\rangle$, with   $ |v(\k + \l/2,-\k + \l/2)|^2 =\delta(\k) \theta(| |\l|^2/4 - k_F^2 | \leq \omega_D)  $,
which only interacts effectively in total momentum variable $\l$, such that equation \eqref{BdGrank1} reads
\begin{multline}\label{highttcT}
 \frac 1g = \int_{ |\l|^2/4 \leq  k_F^2 +\omega_D} \frac{d^3\l}{(2\pi)^3} \frac{\tanh\left (\frac \beta 2\left( (\l/2)^2 - k_F^2\right)\right)}{  \l^2/4  -  k_F^2}.
\end{multline}
Changing variables $\l/2 = \p$ and $\xi = \p^2 - k_F^2$ 
leads to the equation 
$$ 1 = 8 \, \bar g \bar N(0) \int_0^{\omega_D} \frac {d\xi}{\xi} \tanh \frac {\xi}{2T^*},$$ 
with $\bar N(0)$ being the density of states around $|\l| \sim 2 k_F$, and the additional factor $8$ 
steming from the Jakobi-determinant.  
This equation has a solution for temperatures of the form $T^*= \frac {2 e^{\gamma}}{\pi} \omega_D e ^{-1/(8 \bar N(0) \bar g}$. 
If $gN \simeq \bar g \bar N$  we easily see that the temperature which determines the onset of pairs is significantly increased in this case
of an effective interaction depending on the COM-coordinates only. These are now pairs with relativ momentum close to zero and total momenta $|\l| \simeq 2 k_F$.
If, for example, $gN \simeq 0.25$ then the ratio $T^*/T_c$ is of the order of $30$, where $T_c$ is the critical temperature of the classical superconductor.

\subsection{General pairs with $|\p| = |\q|$ and opposite spin} 

Observe that in the limit $\beta \to \infty$ the function $L_\beta$ tends to 
$$  L_\infty (\p,\q) =  \frac{\sgn (\p^2 - k_F^2) + \sgn (\q^2 - k_F^2)}{ \p^2 + \q^2  - 2 k_F^2},$$
which is only non zero for $\{(\p,\q)\, | \, |\p| < k_F, |\q| \leq k_F\}$ and $\{(\p,\q)\, | \, |\p| > k_F, |\q| \geq k_F\}$.

Let us imagine a potential $V$ of the form such that 
$$ |v(\p,\q)|^2 = \delta(|\p| - |\q|) | \tilde v (\p,\q)|^2,$$
then equation \eqref{BdGrank1} turns into
\begin{multline}
\frac 1g = \frac 12 \int_{\R^3} \int_{\R^3} \frac {p^2 d p d\omega_{\p}}{(2\pi)^3}  \frac{q^2 d q d\omega_{\q}}{(2\pi)^3} \delta(p - q)  |\tilde v(p\omega_{ \p}, q \omega_{\q})|^2 \\ \times   \frac{\tanh\left (\frac \beta 2 (p^2 - k_F^2)\right)+ \tanh\left (\frac \beta 2 (q^2 - k_F^2)\right)}{ p^2 + q^2  - 2 k_F^2} \\ =
\frac 12 \int_{\R} p^4 d p f(p)    \frac{\tanh\left (\frac \beta 2 (p^2 - k_F^2)\right)}{ p^2 - k_F^2},
\end{multline}
with  
$$ f(p) = \int_{ {\mathbb{S}}^2 \times {\mathbb{S}}^2} \frac { d\omega_{\p}}{(2\pi)^3}  \frac{d\omega_{\q}}{(2\pi)^3}   |\tilde v(p\omega_{ \p}, p \omega_{\q})|^2  $$ where ${\mathbb{S}}^2$ denotes the unit sphere. 
Here we used the notation  $\omega_{\p} = \frac {\p}{|\p|}$ and $p = |\p|$.
As long as $f(p)$ does not vanish at $p = k_F$ the equation has a solution for an appropriate value of $\beta$. 

\bigskip 
We remark that it is not clear if the temperature $T^*$, which determines the onset of pairing in \eqref{highttcT}  coincides with the critical temperature below which the system displays 
ODLRO, simply because in the BdG equation the direct energy and the Fock term are neglected. 
Due to the importance of the density fluctuations those terms may have an important influence 
on the actual critical temperature $T_c$. In this sense, for an accurate estimate of the critical temperature, one needs to study the full mean field energy functional, or even go beyond. 

\subsection{Conclusion}

We observe that the general BdG-equations allow for the formation of pairs with total momenta different from zero, assuming the existence of effective potentials with attractive center-of-mass contributions. In particular pairs with equal momenta and opposite spin may form at higher critical temperature
compared to classical superconductivity. 

\subsection{Appendix}

In the following we generalize the derivation \cite[Sec. 7]{degennes}  of the linear gap-equation to 
arbitrary two-body interactions $V(\p,\q;\p',\q')$. We do so on a formal basis but anyway use a more mathematical language. 

The BCS-functional for a system  with two-body interaction $V(\p,\q;\p',\q')$, cf. \cite[Equation (2.30)]{HSR} or \cite{bls}, can be written in the concise form
\begin{multline}\label{Fu}
\F(\Gamma) = \Tr \, \epsilon \gamma - T S(\Gamma) \\+\frac 12 \int_{\R^3} \int_{\R^3} \frac {d^3 \p'}{(2\pi)^3} \frac{d^3 \q'}{(2\pi)^3} \overline{\alpha(\p,\q)} V(\p,\q; \p' \q') \,  \alpha(\p',\q') ,
\end{multline}
with $S$ being the entropy $S(\Gamma) = - \Tr \Gamma \log \Gamma$. Further  $\epsilon(\p)  = \p^2 - \k_F^2$, and $\Gamma$ has the form 
$$ \Gamma=  \left(\begin{array}{cc} \gamma & \alpha \\
 \overline{\alpha} &  1 - \overline{\gamma} 
\end{array}\right),$$
with 
$$ \langle a^\dagger_{{\q},\mu} a_{\p,\nu} \rangle = \gamma(\p,\q)\delta_{\nu.\mu} $$ and $\alpha$ defined in \eqref{alph}.  

It is straightforward to see that any minimizer $\Gamma$ of the BCS functional $\F$ has to satisfy the 
equation 
\begin{equation}
\label{BDG-equ} 
\Gamma = \frac 1 {1 + e^{ \beta  H_\Delta}} \,,  \quad  \Delta =  V \alpha,
\end{equation}
where $$\Delta(\p,\q) = \int_{\R^3} \int_{\R^3} \frac {d^3 \p'}{(2\pi)^3} \frac{d^3 \q'}{(2\pi)^3} V(\p,\q; \p' \q') \,  \alpha(\p',\q') ,$$
with 
$$H_\Delta =  \left(\begin{array}{cc} \epsilon & \Delta \\
 \overline{\Delta} & - \overline{\epsilon} 
\end{array}\right).$$

Observe now that equation \eqref{BDG-equ} is an abstract form of the Bogolubov-de-Gennes equations. 
The right hand side only consists of the Cooper-pair wavefunction $\alpha$ only. Hence the equation \eqref{BDG-equ} is determined by the $12$-entry of the  matrix $\Gamma$ only. 
Using further that $1 - \Gamma =  \frac 1 {1 + e^{ - \beta  H_\Delta}}$, we are able to write the equation \eqref{BDG-equ} via the Cauchy integral formula as
\begin{multline} 2 \alpha = - \left[ \tanh \frac \beta 2 H_\Delta \right]_{12} = \\ - \left[ \frac 1 {2\pi i} \int_C \tanh \frac \beta 2 z \frac 1{z- H_0}  \left(\begin{array}{cc} 0 & \Delta \\
 \overline{\Delta} & 0 
\end{array}\right) \frac 1{z - H_{\Delta} } d z \right]_{12} , \end{multline}
where $C$ is an appropriate path around the real axis. 

Next we linearize this equation by replacing $H_\Delta $ by $H_0$ in the resolvent of the right hand side. This leads to the linear equation
\begin{equation}\label{eqalp} 2 \alpha = - L_T \Delta,\end{equation}
where $L_T$ is the two-body operator
\begin{multline}  (L_T  \Delta)(\p,\q)  =  \frac 1{2\pi i} \int_C \tanh \frac \beta 2 z \frac 1{z - \epsilon(\p)} \Delta(\p,\q) \frac 1{z + \overline{ \epsilon(\q)} }  d z  \\=  \frac{\tanh\left( \frac \beta 2 \epsilon(\p) \right)+\tanh\left( \frac \beta 2 \epsilon(\q) \right)}{\epsilon(\p) + \epsilon(\q)} \Delta(\p,\q).
\end{multline}
Applying $V$ to both sides of the equation \eqref{eqalp} one obtains equation \eqref{BdG} .
We remark that in case the potential $V$ is periodic in the COM-variables the functional $\F$ is a-priori ill defined. However this is easily repaired 
by interpreting the trace in \eqref{Fu} as trace per unit volume as in \cite{FHSS,FHSS2,HSR}.

\subsection{Acknowledgement} The work was partially supported by U.S. National Science Foundation Grant DMS 1301555 and the German Humboldt
Foundation. M. L. would like to thank the Department of Mathematics at the University of T\"ubingen for their hospitality. 




\end{document}